\documentclass[runningheads]{llncs}
\usepackage[T1]{fontenc}
\usepackage{graphicx}
\usepackage{amsmath}
\usepackage{amsfonts}
\usepackage{amssymb}
\usepackage{amsbsy}
\usepackage{subfigure}

\usepackage[pdftex,colorlinks=true,urlcolor=blue,citecolor=black,anchorcolor=black,linkcolor=black,bookmarks=false]{hyperref}

\usepackage{xcolor}

\usepackage{listings}
\usepackage{xfrac}
\def\onehalf{\sfrac 1 2}

\usepackage{array}
\newcolumntype{F}{>{\centering\arraybackslash}p{3.8cm}}
\newcolumntype{G}{>{\centering\arraybackslash}p{6.7cm}}
\newcolumntype{d}[1]{D{.}{.}{#1}}

\usepackage{tikz}
\tikzset{font={\fontsize{10pt}{12}\selectfont}}
\usetikzlibrary{3d, decorations.text, shapes.arrows, positioning, fit, backgrounds}
\usetikzlibrary{shapes, calc, arrows, decorations.markings}

\newcommand{\B}{\mathbb B}
\newcommand{\Hd}{\{0,1,*\}}
\newcommand{\range}[1]{\{1, \ldots, #1\}}
\newcommand{\sync}{\mathrm{s}}
\newcommand{\fa}{\mathrm{a}}
\newcommand{\ga}{\mathrm{g}}
\newcommand{\MP}{\mathrm{MP}}
\newcommand{\rank}{\operatorname{rank}}
\newcommand{\confs}{\operatorname{c}}
\newcommand{\myh}[2]{h^{\langle #1,#2\rangle}}
\newcommand{\tr}{\operatorname{tr}}
\newcommand{\exone}{\text{(A)}}
\newcommand{\extwo}{\text{(B)}}

\newcommand{\myscale}{1}

\tikzstyle{every matrix}=[ampersand replacement=\&]
\tikzstyle{shorthandoff}=[]
\tikzstyle{shorthandon}=[]

\begin{document}
\title{Variable-Depth Simulation of Most Permissive Boolean Networks}
%
%
\author{%
Théo Roncalli\inst{1}
\and
Loïc Paulevé\inst{1}\orcidID{0000-0002-7219-2027}}
\authorrunning{T. Roncalli and L. Paulevé}
\institute{
Univ. Bordeaux, CNRS, Bordeaux INP, LaBRI, UMR 5800\\
F-33400 Talence, France\\
\email{theo.roncalli@universite-paris-saclay.fr}\\
\email{loic.pauleve@labri.fr}}
\maketitle
\begin{abstract}
In systems biology, Boolean networks (BNs) aim at modeling the qualitative dynamics of quantitative
biological systems.
Contrary to their (a)synchronous interpretations, the Most Permissive (MP) interpretation guarantees
capturing all the trajectories of any quantitative system compatible with the BN, without additional
parameters.
Notably, the MP mode has the ability to capture transitions related to the heterogeneity of time
scales and concentration scales in the abstracted quantitative system and which are not captured by
asynchronous modes.
So far, the analysis of MPBNs has focused on Boolean dynamical properties, such as the existence of
particular trajectories or attractors.

This paper addresses the sampling of trajectories from MPBNs in order to quantify the propensities
of attractors reachable from a given initial BN configuration.
The computation of MP transitions from a configuration is performed by iteratively discovering
possible state changes.
The number of iterations is referred to as the \emph{permissive depth}, where the first depth
corresponds to the asynchronous transitions.
This permissive depth reflects the potential concentration and time scales heterogeneity along the
abstracted quantitative process.
The simulation of MPBNs is illustrated on several models from the literature, on which the depth
parametrization can help to assess the robustness of predictions on attractor propensities changes
triggered by model perturbations.
\end{abstract}

\section{Introduction}
\label{sec:intro}

Boolean networks (BNs) have been employed to model the temporal evolution of gene expression and
protein activities in biological systems~\cite{kauffman69,thomas73,Zanudo21,Montagud22}.
A BN is composed of a finite set of components having two states, either 0 (false) or 1
(true).
The BN then specifies in which contexts the components can change state, such as
``component $3$ can switch to state 1 if and only if component $1$ is in state 0 \emph{and}
component $2$ is in state 1; in all other contexts, component $3$ can only switch to 0''.
These rules can be given as one Boolean function per component, associating each possible
configuration (which associates each component to a state) to a Boolean value:
in our example, the function of component $3$ is $f_3: \{0,1\}^n \to \{0,1\}$ with $f_3(x) =
\neg x_1\wedge x_2$, where $\neg$ and $\wedge$ denote the logical negation and conjunction,
respectively.

The \emph{execution} of a BN relies on a given \emph{update mode} which specifies how are
computed the evolution of the states of components.
In the systems biology literature, two main update modes are widely employed: the \emph{synchronous}
(or \emph{parallel}) and \emph{fully-asynchronous}.
In synchronous, each component is updated simultaneously: the configuration $x=(x_1, x_2, x_3)$ evolves
in one step to $(f_1(x), f_2(x), f_3(x))$.
In fully-asynchronous, only one component can be updated at a time, possibly leading to
non-deterministic transitions: the configuration $x$ can evolve either to $(f_1(x), x_2, x_3)$,
$(x_1, f_2(x), x_3)$, or $(x_1, x_2, f_3(x))$.
Whatever the chosen update mode, as there is a finite number of configurations, an execution
eventually reaches a set of \emph{limit} configurations which will be visited infinitely often.
These sets of configurations are called \emph{attractors}: each execution eventually reaches and
stays within one attractor.
Attractors are prominent dynamical features when modeling biological processes:
they represent stable behaviors and are usually associated to cellular phenotypes.

Computing the attractors \emph{reachable} from a given initial configuration is at the core of many
studies of biological processes with BNs~\cite{CellDeath2010,TumorInvasion,Bladder2015,Mendes2018,Montagud22}.
These studies typically involve comparing the effect of a network mutation (forcing some components to have
fixed value) on the sets of reachable attractors and their propensities.
This later notion is often related to the number of paths leading from the initial configuration to
each attractor: under some mutations, the same set of attractors may still be reachable, but the
proportion of trajectories leading to them may substantially differ.
This motivated the development of \emph{simulation} algorithms for BNs in order to sample
trajectories and quantify the propensities to reach attractors.
These methods replace the non-determinism of asynchronous transitions by a probabilistic
choice~\cite{Mendes2018,Stoll2012}.

Whenever employed as qualitative models of quantitative systems,
dynamics of BNs aim at giving a coarse-grained view of system dynamics without requiring numerous
quantitative parameters.
However, the Boolean (a)synchronous modes are not correct abstractions of quantitative
dynamics~\cite{MPBNs}: they lead at the same time to predict spurious transitions and, importantly,
preclude transitions which are actually possible when considering delays for instance.
Let us illustrate this with the following BN, denoted $\eqref{eq:example1}$ in the rest of the text:
\begin{equation}\label{eq:example1}
    f_1(x) = 1 \nonumber \qquad
    f_2(x) = x_1 \nonumber \qquad
    f_3(x) = (\neg x_1 \wedge x_2) \vee x_3 \nonumber
    \tag{A}
\end{equation}
From the initial configuration where all components are 0, we write $000$, there is only one possible
transition: the activation of $1$, leading to the configuration $100$. Then, again, only one
transition is possible, activating $2$, thus leading to $110$. There, the execution will stay
infinitely on this configuration: no other state changes are possible: from $000$ it is impossible
to eventually activate $3$, and $\{110\}$ is the only reachable attractor.
However, it is known that this system can actually activate $3$ for a range of kinetics,
as observed experimentally~\cite{Schaerli2014}, and easily captured with quantitative
models~\cite{Ishihara2005,Rodrigo2011}.
Indeed, consider that $1$ has actually several activation levels: one intermediate $\onehalf$
which is sufficient to activate $2$ but not enough sufficient to inhibit $3$: when in
$\onehalf 00$, $2$ can be activated, going to $\onehalf 10$, and then $3$ can change state,
going to $\onehalf 11$.
When $1$ becomes fully active ($111$), $3$ will self-maintain its activation.
Thus, a correct Boolean analysis of the BN $f$ above should conclude that from $000$, two attractors
are reachable: $\{110\}$, when $1$ goes rapidly to its maximum level, and $\{111\}$ when $3$ had
time to activate before the full activation of $1$.
This example shows the limit of (a)synchronous interpretations of BNs when used as abstraction
of quantitative systems: they enforce that the Boolean 0 matches with the quantitative 0, and the
Boolean 1 matches with the quantitative non-zero ($>0$), making impossible to capture transitions happening
at different activation levels or different time scales.

The \emph{Most Permissive} (MP) update mode of BNs~\cite{MPBNs,automata21} is a recently-introduced
execution paradigm which enables capturing dynamics precluded by the asynchronous ones.
The main idea behind the MP update mode is to systematically consider a potential delay when a
component changes state, and consider any additional transitions that could occur if the changing
component is in an intermediate state.
It can be modeled as additional \emph{dynamic} states ``increase'' ($\nearrow$) and ``decrease''
($\searrow$): when a component can be activated, it will first go through the ``increase'' state
where it can be interpreted as either 0 or 1 by the other components, until eventually reaching the
Boolean 1 state.
With the previous example, starting from $000$, the first component is put as increasing, thus going
to the MP configuration $\nearrow\!00$. In this configuration, $2$ can be activated because $\nearrow$ can
be interpreted as 1, leading to $\nearrow\nearrow\!0$. Then, $3$ can be activated, as it can
interpret the dynamic state of $1$ as the Boolean $0$, and $2$ as the Boolean $1$.
This model the fact that the component $1$ is not high enough for inhibiting $3$, while $2$ is high
enough to activate it. Thus, in this example, there is a trajectory from $000$ to $111$, i.e., a
configuration where all the components are active.
As $111$ is a fixed point of $f$, the MP analysis would thus conclude that two attractors are
reachable from $000$: $\{110\}$ and $\{111\}$.

The MP update mode brings a formal abstraction property
to BN dynamics with respect to quantitative models, without requiring additional parameters:
essentially, MPBNs capture any behavior that is achievable by any quantitative
model being compatible with the logic of the BN.
This include, for instance, models which result from introducing quantitative parameters, such as
transition speed and interaction thresholds.
We give here a brief informal overview of the property (see~\cite{MPBNs} for details):
let us consider multivalued networks (MNs) of dimension $n$ where components can have values in
$\mathbb M = \{0, \dots, m\}$ for some $m\in\mathbb N_{>0}$.
A MN can be considered as map from multivalued configurations to the derivative of the value of the
components, i.e., of the form $F:\mathbb M^n\to \{-1,0,1\}$.
A multivalued configuration $z\in \mathbb M^n$ can be \emph{binarized} by associating components
with value 0 to the Boolean state 0, components with value $m$ to the Boolean state 1, and other
components to any Boolean state.
A MN $F$ is a \emph{refinement} of a BN $f$ whenever for any multivalued configuration $z$ and for
each component $i$, if component $i$ increases (resp. decreases), i.e., $F_i(z) > 0$ (resp. $F_i(z)
< 0$), there exists a binarization of this configuration  such that $f_i$ is evaluated to 1
(resp. to 0).
Then, for any pair of Boolean configurations $x$ and $y$, if there exists a trajectory from $m\cdot x$
to $m\cdot y$ in the asynchronous dynamics of the MN, there necessarily exists a trajectory from $x$ to $y$ in the MP dynamics of the BN.
Moreover, to any transition computed according the MP update mode, there is refinement of the BN which realizes it.
One of the major consequences of the abstraction property is that if a configuration is not reachable
from another with the MP update mode, then no quantitative models being compatible with the BN
can produce the trajectory.
In addition, the MP mode has a lower computational complexity for computing reachability and attractor
properties, enabling formal analysis of genome-scale BNs~\cite{MPBNs,VLBNs}.

However, the simulation of MPBNs, i.e., the sampling of transitions following the MP update mode,
has not been addressed so far.
Thus, besides computing Boolean properties, such as the existence or absence of reachable attractors, there is no
algorithm nor tools to approach the effect of a mutation on the propensities to reach attractors,
as we mentioned above with asynchronous BNs.

In this paper, we present a first algorithm for sampling trajectories of BNs with the MP update
mode, subject to additional simulation parameters for assigning probabilities to the transitions.
The MP transitions enabled from a single configuration are computed iteratively, and we refer to the
number of times this iteration is performed as the \emph{depth} of the MP computation.
At depth 1, the transitions match with the asynchronous update mode, but further depths bring
additional behaviors.
These iterations capture possibly
different time scales: while component $1$ is changing (depth 1), $2$ can change (depth 2); then while $2$ is changing,
$3$ can change (depth 3), etc.
In our simulation algorithm, the probability of a transition can be affected by its depth and
the number of components it changes simultaneously.
Thus, similarly to~\cite{Mendes2018,Stoll2012} with the fully-asynchronous mode, our sampling of
trajectories can be assimilated to a random walk in the MP dynamics, where the
probability of transitions can be tuned with the simulation parameters.
As we will show on case studies, the MP interpretation can lead to drastic changes in the predicted
probabilities of reachable attractors, enabling assessing the robustness of prediction to the
heterogeneity of time scales and concentration scales in the quantitative system captured by the
discrete MP dynamics.
Nevertheless, the simulation parameters and derived transition probabilities are
empirical and cannot be formally related to a putative abstracted system.

\section{Background}
\label{sec:defs}

\subsection{Boolean networks and dynamics}

A \emph{Boolean network} (BN) of dimension $n$ is a function $f:\B^n\to \B^n$,
with $\B=\{0,1\}$.
For each $i\in\range n$, $f_i:\B^n\to \B$ is the \emph{local function} of component $i$.
The Boolean vectors $x\in \B^n$ are the \emph{configurations} of $f$, where for each
$i\in\range n$, $x_i$ is the state of $i$.
Given two configurations $x,y\in\B^n$, the components having a different state are noted
$\Delta(x,y)=\{i\in\range n\mid x_i\neq y_i\}$.

The \emph{influence graph} $G(f)$ of a BN $f$ is a signed digraph whose nodes are the components
and edges mark the dependencies between them in the local functions:
for all $i,j\in\range n$, there is an edge $i\xrightarrow{s} j$ in $G(f)$ with $s\in\{-1,+1\}$ if and
only if there exists a configuration $x\in\B^n$ with $x_i=0$ such that
$s = f_j(x_1,\ldots,x_{i-1},1,x_{i+1},\ldots,x_n) - f_j(x)$:
the sole increasing of component $i$ causes $f_j$ to increase ($s=+1$) or decrease ($s=-1$).
Note there may exist $i$ and $j$ so that both $i \xrightarrow{+1} j$ and $i\xrightarrow{-1}j$
are in $G(f)$, for instance with $f_j(x) = x_i \operatorname{~xor~} x_k$.
A BN $f$ is \emph{locally monotone} whenever there is no $i,j\in\range n$ such that both
$i \xrightarrow{+1} j$ and $i\xrightarrow{-1}j$ are in $G(f)$, i.e.,
each of its local function is \emph{unate}.
Computing $G(f)$ is a DP-complete problem (both in NP and coNP)~\cite{Crama2011}.
For the example BN $f$ of Eq.\eqref{eq:example1}, $G(f)= \{ 1\xrightarrow{+1} 2, 1\xrightarrow{-1} 3, 2
\xrightarrow{+1} 3, 3\xrightarrow{+1} 3\}$: it is locally monotone.

An \emph{update mode} $\mu$ of $f$ specifies a binary transition relation between configurations
$\to_\mu\,\subseteq\B^n\times\B^n$. Classical update modes include the \emph{synchronous} (or \emph{parallel}) update mode
where \(x\to_{\sync} y\) iff \(x\neq y\) and \(y=f(x) \);
the \emph{fully-asynchronous} update mode
where \(x\to_{\fa} y\) iff $x$ and $y$ differ on only one component $i$ and \(y_i=f_i(x)\);
and the (general) \emph{asynchronous} update mode where
\(x\to_{\ga} y\) iff \(x\neq y\) and for each component \(i\in\Delta(x,y), y_i=f_i(x)\).

Given an update mode $\mu$, a configuration $y\in\B^n$ is \emph{reachable} from a configuration
$x\in\B^n$, noted $x\to_\mu^* y$, if and only if either $x=y$ or there exists a sequence of transitions $x\to_\mu \cdots\to_\mu y$.
A non-empty set of configurations $A\subseteq \B^n$ is an \emph{attractor} if and only if
for each pair of configurations $x,y\in A$, $y$ is reachable from $x$, and there is no configuration
$z\in\B^n\setminus A$ that is reachable by a configuration in $A$.
Remark that attractors are the bottom strongly connected components of the digraph $(\B^n,\to_\mu)$.
Whenever $A$ is a singleton configuration $\{x\}$, it is said to be a \emph{fixed point} of the
dynamics. Otherwise, $A$ is a cyclic attractor.
In the case of (a)synchronous and MP update modes, the fixed points of the dynamics match exactly
with the fixed points of $f$, i.e., the configurations $x\in\B^n$ such that $f(x)=x$.
Finally, the \emph{strong basin} of an attractor $A$ is the set of configurations that can reach $A$
and no other distinct attractor.

Fig.~\ref{fig:inputmodule} shows the transitions computed with the above defined update
modes with the BN $f$ defined as follows:
\begin{equation}
    f_1(x) = x_{1} \wedge \neg x_{3}
    \qquad
    f_2(x) = x_1
    \qquad
    f_3(x) = \neg x_1
\tag{B}
\label{ex:inputmodule}
\end{equation}
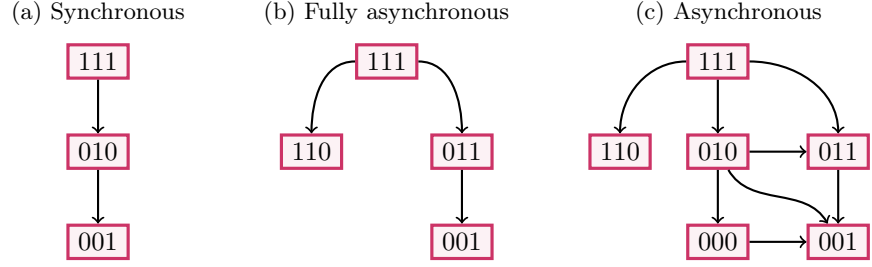
\begin{figure}[t]

    \centering

    \begin{tabular}{ccccc}
    \vspace*{0.2cm}
    (a) Synchronous
    &
    \hspace*{0.85cm}
    &
    (b) Fully asynchronous
    &
    \hspace*{0.85cm}
    &
    (c) Asynchronous
    \\

    \begin{tikzpicture}[every text node part/.style={align=center}]

    \node[minimum width=0.80cm, draw, rectangle, color=purple!80, text=black!100, fill=purple!5, very thick] (1) at (0,0) {$111$};
    \node[minimum width=0.80cm, draw, rectangle, color=purple!80, text=black!100, fill=purple!5, very thick] (2) at (0,-1.2) {$010$};
    \node[minimum width=0.80cm, draw, rectangle, color=purple!80, text=black!100, fill=purple!5, very thick] (3) at (0,-2.4) {$001$};

    \draw[thick,->] (1) to [out=270,in=90] (2);
    \draw[thick,->] (2) to [out=270,in=90] (3);

    \node at (0,-3.3) {$\quad$};

    \end{tikzpicture}

    &

    &

    \begin{tikzpicture}[every text node part/.style={align=center}]

    \node[minimum width=0.80cm, draw, rectangle, color=purple!80, text=black!100, fill=purple!5, very thick] (1) at (0,0) {$111$};
    \node[minimum width=0.80cm, draw, rectangle, color=purple!80, text=black!100, fill=purple!5, very thick] (2) at (-1.0,-1.2) {$110$};
    \node[minimum width=0.80cm, draw, rectangle, color=purple!80, text=black!100, fill=purple!5, very thick] (3) at (1.0,-1.2) {$011$};
    \node[minimum width=0.80cm, draw, rectangle, color=purple!80, text=black!100, fill=purple!5, very thick] (4) at (1.0,-2.4) {$001$};

    \draw[thick,->] (1) to [out=180,in=90] (2);
    \draw[thick,->] (1) to [out=0,in=90] (3);
    \draw[thick,->] (3) to [out=270,in=90] (4);

    \node at (0,-3.3) {$\quad$};

    \end{tikzpicture}

    &

    &

    \begin{tikzpicture}[every text node part/.style={align=center}]

        \node[minimum width=0.80cm, draw, rectangle, color=purple!80, text=black!100, fill=purple!5, very thick] (1) at (0.0,0.0) {$111$};
        \node[minimum width=0.80cm, draw, rectangle, color=purple!80, text=black!100, fill=purple!5, very thick] (2) at (-1.3,-1.2) {$110$};
        \node[minimum width=0.80cm, draw, rectangle, color=purple!80, text=black!100, fill=purple!5,
            very thick] (3) at (1.6,-1.2) {$011$};
        \node[minimum width=0.80cm, draw, rectangle, color=purple!80, text=black!100, fill=purple!5,
        very thick] (4) at (1.6,-2.4) {$001$};
        \node[minimum width=0.80cm, draw, rectangle, color=purple!80, text=black!100, fill=purple!5, very thick] (5) at (0.0,-1.2) {$010$};
        \node[minimum width=0.80cm, draw, rectangle, color=purple!80, text=black!100, fill=purple!5,
        very thick] (c000) at (0,-2.4) {$000$};

        \draw[thick,->] (1) to [out=180,in=90] (2);
        \draw[thick,->] (1) to [out=270,in=90] (5);
        \draw[thick,->] (1) to [out=0,in=90]   (3);
        \draw[thick,->] (3) to [out=270,in=90] (4);
        \draw[thick,->] (5) to [out=300,in=120] (4);
        \draw[thick,->] (5) to [out=270,in=90] (c000);
        \draw[thick,->] (c000) to [out=0,in=180]   (4);
        \draw[thick,->] (5) to [out=0,in=180]   (3);

        \node at (0,-3.3) {$\quad$};

    \end{tikzpicture}
    \\

    \end{tabular}

    \vspace*{-0.5cm}
    \caption{Transitions of the BN \eqref{ex:inputmodule} from the configuration $111$ with different modes}
    \label{fig:inputmodule}

\end{figure}
This model has two attractors, being fixed points $001$ and $110$.
With the synchronous mode, only $001$ is reachable from $111$, whereas both are reachable with the
asynchronous modes.

\subsection{Sub-hypercubes and closures}

The MP update mode and the algorithm presented in this paper gravitate around the notion of
sub-hypercube of $\B^n$ and their partial closure by $f$.

A Boolean \emph{sub-hypercube} of dimension $n$ is specified by a vector in $\Hd^n$, where
components having value $*$ are said \emph{free}, otherwise they are \emph{fixed}.
The number of free components in a sub-hypercube $h\in\Hd^n$ is denoted by
$\rank(h)=|\{i\in \range n\mid h_i=*\}|$.
A sub-hypercube $h$ has $2^{\rank(h)}$ vertices, denoted by
$\confs(h)=\{x\in\B^n\mid\forall i\in\range n,\, h_i\in\B\Rightarrow x_i=h_i\}$.
A sub-hypercube $h'\in\Hd^n$ is \emph{smaller} than a sub-hypercube $h\in\Hd^n$ whenever for each
$i\in\range n$ fixed in $h$ ($h_i\in\B$), it is fixed to the value in $h'$ ($h'_i=h_i)$.
Then, remark that $\confs(h')\subseteq \confs(h)$.

A sub-hypercube $h$ is \emph{closed} by a BN $f$ if the result of $f$ applied to any of its vertices
is one of its vertices:
$\forall x\in\confs(h),\, f(x)\in\confs(h)$;
it is also known as a \emph{trap space}.
Given components $K\subseteq \range n$, $h$ is \emph{$K$-closed} by $f$ whenever,
for each component $i\in K$, either $i$ is free in $h$, or $f_i$ applied on any vertices of $h$
results in the fixed value $h_i$. In other words, for all configurations in the $K$-closed sub-hypercube $h$,
the next states of the components $i \in K$ are in $h$:
\begin{equation}
    \forall x\in\confs(h),\, \forall i\in K,\, h_i\neq *\Rightarrow f_i(x)=h_i.
    \label{eq:k-closed}
\end{equation}

Let us consider the BN $f$ defined in $\extwo$.
The sub-hypercube $h = 0**$ is closed by $f$, where $c(0**) = \{000,010,001,011\}$.
This trap space indicates that component $1$ can never get activated once deactivated.
The sub-hypercube $h = 0*1$ is $\{1,2\}$-closed by $f$ as
$f_{1}\left(001\right) = f_{1}\left(011\right) = 0$.
Sub-hypercubes $1*0$ and $**0$ are also $\{1,2\}$-closed by $f$,
where $1*0$ is smaller than $**0$.
In contrast, $1*1$ is not $\{1,2\}$-closed by $f$.

\subsection{The Most Permissive update mode}

Whenever a component changes state, e.g., increases from its minimal value $0$ to its maximal value
$1$, the MP mode captures any behavior that could arise in the course of this change.
For example, it may be that at some point, the component becomes high enough to activate one of its
targets, whereas it remains not high enough to activate another one (because it has not reached its
maximal value yet).
These \emph{dynamic states} can be captured by sub-hypercubes, where the changing components are
free: they can be read both as 0 or 1.

Formally, the MP update mode can be defined as follows.
Given a set of components $K\subseteq\range n$, let us denote by $\myh x K$ the \emph{smallest}
sub-hypercube of dimension $n$ that contains $x$ and that is $K$-closed by $f$.
There is an MP transition from $x$ to $y$ whenever (a) $y$ is a vertex of $\myh x K$, and (b) the new
state of all the components in $K$ can be computed from $\myh x K$:
\begin{equation}
    \label{eq:mp}
    \begin{split}
        \forall x,y\in\B^n,\quad x\to_{\MP} y \Longleftrightarrow \exists& K \subseteq \range n:
    y\in\confs(\myh{x}{K})
    \\
                                                                         &\wedge \forall i\in K, \exists z\in\confs(\myh{x}{K}): y_i=f_i(z)\enspace.
    \end{split}
\end{equation}
The abstraction properties and complexity results are detailed in~\cite{MPBNs}.
One can remark that the transition and reachability relations are identical:
$y$ is reachable from $x$ if and only if $x\to_{\MP} y$.

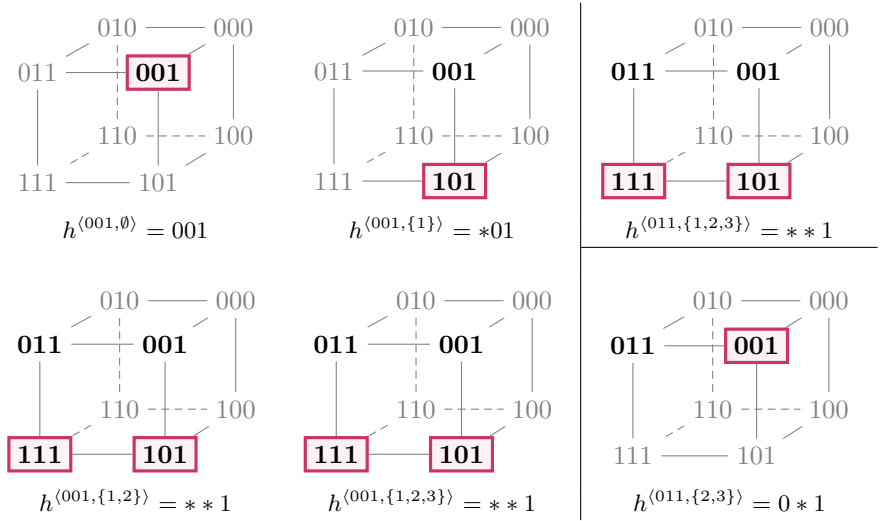
\begin{figure}[bt]
    \centering
    \vspace*{0.3cm}

    \begin{tabular}{cc|c}

    \scalebox{\myscale}{
    
        \begin{tikzpicture}
            \matrix[column sep=0.8cm,row sep=1cm,gray]{
                \node (s011) {$011$};\&
                \node[draw, rectangle, color=purple!80, text=black!100, fill=purple!5, very thick] (s001) {$\mathbf{001}$};
                \\
                \node (s111) {$111$}; \&
                \node (s101) {$101$};
                \\
            };
            
            \matrix[column sep=0.8cm,row sep=1cm,shift={(1cm,0.6cm)},gray] {
                \node (s010) {$010$}; \&
                \node (s000) {$000$};
                \\
                \node (s110) {$110$}; \&
                \node (s100) {$100$};
                \\
            };
            
            \path[gray]
            
            (s000) edge (s010) edge (s100) edge (s001)
            (s110) edge[densely dashed] (s010) edge[densely dashed] (s100) edge[densely dashed] (s111)
            (s001) edge (s011) edge (s101)
            (s111) edge (s011) edge (s101)
            (s010) edge (s011)
            (s100) edge (s101)
            ;

        \end{tikzpicture}
    }
    
    &
    
    \scalebox{\myscale}{
    
        \begin{tikzpicture}
            \matrix[column sep=0.8cm,row sep=1cm,gray]{
                \node (s011) {$011$};\&
                \node[rectangle, text=black!100] (s001) {$\mathbf{001}$};
                \\
                \node (s111) {$111$}; \&
                \node[draw, rectangle, color=purple!80, text=black!100, fill=purple!5, very thick] (s101) {$\mathbf{101}$};
                \\
            };
            
            \matrix[column sep=0.8cm,row sep=1cm,shift={(1cm,0.6cm)},gray] {
                \node (s010) {$010$}; \&
                \node (s000) {$000$};
                \\
                \node (s110) {$110$}; \&
                \node (s100) {$100$};
                \\
            };
            
            \path[gray]
            
            (s000) edge (s010) edge (s100) edge (s001)
            (s110) edge[densely dashed] (s010) edge[densely dashed] (s100) edge[densely dashed] (s111)
            (s001) edge (s011) edge (s101)
            (s111) edge (s011) edge (s101)
            (s010) edge (s011)
            (s100) edge (s101)
            ;

        \end{tikzpicture}
    }
    
    &

    \scalebox{\myscale}{
    
        \begin{tikzpicture}
            \matrix[column sep=0.8cm,row sep=1cm,gray]{
                \node[rectangle, text=black!100] (s011) {$\mathbf{011}$};\&
                \node[rectangle, text=black!100] (s001) {$\mathbf{001}$};
                \\
                \node[draw, rectangle, color=purple!80, text=black!100, fill=purple!5, very thick] (s111) {$\mathbf{111}$}; \&
                \node[draw, rectangle, color=purple!80, text=black!100, fill=purple!5, very thick] (s101) {$\mathbf{101}$};
                \\
            };
            
            \matrix[column sep=0.8cm,row sep=1cm,shift={(1cm,0.6cm)},gray] {
                \node (s010) {$010$}; \&
                \node (s000) {$000$};
                \\
                \node (s110) {$110$}; \&
                \node (s100) {$100$};
                \\
            };
            
            \path[gray]
            
            (s000) edge (s010) edge (s100) edge (s001)
            (s110) edge[densely dashed] (s010) edge[densely dashed] (s100) edge[densely dashed] (s111)
            (s001) edge (s011) edge (s101)
            (s111) edge (s011) edge (s101)
            (s010) edge (s011)
            (s100) edge (s101)
            ;

        \end{tikzpicture}
    }
    
    \\
    \(\myh {001} {\emptyset} = 001\)
    &
    \(\myh {001} {\{1\}} = *01\)
    &
    \(\myh {011} {\{1,2,3\}} = **1\)
 
    \\\cline{3-3}

    &&

    \\
    
    \scalebox{\myscale}{
    
        \begin{tikzpicture}
            \matrix[column sep=0.8cm,row sep=1cm,gray]{
                \node[rectangle, text=black!100] (s011) {$\mathbf{011}$};\&
                \node[rectangle, text=black!100] (s001) {$\mathbf{001}$};
                \\
                \node[draw, rectangle, color=purple!80, text=black!100, fill=purple!5, very thick] (s111) {$\mathbf{111}$}; \&
                \node[draw, rectangle, color=purple!80, text=black!100, fill=purple!5, very thick] (s101) {$\mathbf{101}$};
                \\
            };
            
            \matrix[column sep=0.8cm,row sep=1cm,shift={(1cm,0.6cm)},gray] {
                \node (s010) {$010$}; \&
                \node (s000) {$000$};
                \\
                \node (s110) {$110$}; \&
                \node (s100) {$100$};
                \\
            };
            
            \path[gray]
            
            (s000) edge (s010) edge (s100) edge (s001)
            (s110) edge[densely dashed] (s010) edge[densely dashed] (s100) edge[densely dashed] (s111)
            (s001) edge (s011) edge (s101)
            (s111) edge (s011) edge (s101)
            (s010) edge (s011)
            (s100) edge (s101)
            ;

        \end{tikzpicture}
    }
    
    &

    \scalebox{\myscale}{
    
        \begin{tikzpicture}
            \matrix[column sep=0.8cm,row sep=1cm,gray]{
                \node[rectangle, text=black!100] (s011) {$\mathbf{011}$};\&
                \node[rectangle, text=black!100] (s001) {$\mathbf{001}$};
                \\
                \node[draw, rectangle, color=purple!80, text=black!100, fill=purple!5, very thick] (s111) {$\mathbf{111}$}; \&
                \node[draw, rectangle, color=purple!80, text=black!100, fill=purple!5, very thick] (s101) {$\mathbf{101}$};
                \\
            };
            
            \matrix[column sep=0.8cm,row sep=1cm,shift={(1cm,0.6cm)},gray] {
                \node (s010) {$010$}; \&
                \node (s000) {$000$};
                \\
                \node (s110) {$110$}; \&
                \node (s100) {$100$};
                \\
            };
            
            \path[gray]
            
            (s000) edge (s010) edge (s100) edge (s001)
            (s110) edge[densely dashed] (s010) edge[densely dashed] (s100) edge[densely dashed] (s111)
            (s001) edge (s011) edge (s101)
            (s111) edge (s011) edge (s101)
            (s010) edge (s011)
            (s100) edge (s101)
            ;

        \end{tikzpicture}
    }

    &
    
    \scalebox{\myscale}{
    
        \begin{tikzpicture}
            \matrix[column sep=0.8cm,row sep=1cm,gray]{
                \node[rectangle, text=black!100] (s011) {$\mathbf{011}$};\&
                \node[draw, rectangle, color=purple!80, text=black!100, fill=purple!5, very thick] (s001) {$\mathbf{001}$};
                \\
                \node (s111) {$111$}; \&
                \node (s101) {$101$};
                \\
            };
            
            \matrix[column sep=0.8cm,row sep=1cm,shift={(1cm,0.6cm)},gray] {
                \node (s010) {$010$}; \&
                \node (s000) {$000$};
                \\
                \node (s110) {$110$}; \&
                \node (s100) {$100$};
                \\
            };
            
            \path[gray]
            
            (s000) edge (s010) edge (s100) edge (s001)
            (s110) edge[densely dashed] (s010) edge[densely dashed] (s100) edge[densely dashed] (s111)
            (s001) edge (s011) edge (s101)
            (s111) edge (s011) edge (s101)
            (s010) edge (s011)
            (s100) edge (s101)
            ;

        \end{tikzpicture}
    }
 
    \\
    \(\myh {001} {\{1,2\}} = **1\)
    &
    \(\myh {001} {\{1,2,3\}} = **1\)
    &
    \(\myh {011} {\{2,3\}} = 0*1\)
    \\

    \end{tabular}
    \caption{Computation of sub-hypercubes with BN $f(x) = (1, x_1, (\neg x_1 \wedge x_2) \vee x_3)$}
    \label{fig:hypercubes}
\end{figure}

As this formalism serves as the cornerstone for the simulation algorithm, let us illustrate
it by using the BN~\eqref{eq:example1}.
Fig.~\ref{fig:hypercubes} depicts the smallest sub-hypercubes with respect to a configuration $x$ and a set $K$.
Bold configurations belong to $\myh x K$ and those satisfying reachability condition (b) are boxed.
The left side focuses on the initial configuration $001$ and shows an iterative computation of $\myh {001}
K$ from $K=\emptyset$ to $K=\{1,2,3\}$.
As component $1$ can flip in configuration $001$, one obtains $\myh {001} {\{1\}} = *01$.
If component $1$ is active, component $2$ can flip, leading to $\myh {001} {\{1,2\}} = **1$.
For all configurations $z \in c\left(\myh {001} {\{1,2\}}\right)$, component $3$ stays active: $f_{3}\left(z\right) = 1$.
Hence, one obtains $\myh {001} {\{1,2,3\}} = **1$.
Here, some reachable configurations from $x$ can be deduced.
For instance, transitions ${001} \to_{\MP} {101}$ and ${001} \to_{\MP} {111}$ exist because reachability properties (a) and (b) are verified,
However, ${000}$ and $011$ do not verify properties (a) and (b) respectively.
Multiple sub-hypercubes may be required to capture all the MP transitions.
This is illustrated in Fig.~\ref{fig:hypercubes} on the right side with the configuration $011$.
The sub-hypercube $\myh {011} {\{1,2,3\}}$ does not capture
the transition to $001$ because property (b) is not satisfied, while $\myh {011} {\{2,3\}}$ does.

\section{Simulation algorithm}

\subsection{Main principle}

In its basic formal definition~\eqref{eq:mp}, the MP update mode defines the possible next configurations by
considering all the subsets of components $K\subseteq \range n$ and compute the associated
sub-hypercube $\myh x K$, that is the smallest $K$-closed sub-hypercube containing $x$.
However, part of these subsets may be redundant.
Let $H$ be the components free in $\myh x K$~\eqref{eq:H}.
By the minimality of $\myh x K$, it means that for each component $i\in H$, there exists a
configuration $y\in c(\myh x K)$ such that $f_i(y)\neq x_i$.
The set $H$ can then be split in two:
the components $J$ for which their local function can be evaluated both to 0 and 1 from the
configurations of $\myh x K$~\eqref{eq:J},
and the components $L$ for which their local function is always evaluated to $\neg x_i$ from any of
the configurations of $\myh x K$:
\begin{align}
    H &= \left\{ i\in K\mid \myh x K_i=*\right\}\enspace,\label{eq:H}\\
    J &= \left\{ i\in H\mid \exists y,z\in\confs(\myh x K): f_i(y)=0\land f_i(z)=1 \right\}\enspace,\label{eq:J} \\
    L &= \left\{ i\in H\mid \forall y\in\confs(\myh x K): f_i(y)\neq x_i\right\}\enspace.\label{eq:L}
\end{align}
We have $H = J\cup L$ and $J\cap L=\emptyset$.
Then, remark that in~\eqref{eq:mp}, for a fixed $x$ and $K$, there exists an MP transition
$x\to_{\MP} y$ if and only if $L\subseteq \Delta(x, y)\subseteq H$.
Indeed, for each component $i\in\range n$, if $y_i\neq x_i$ then necessarily $i\in H$.
Moreover, if $i\in L$, the second condition of~\eqref{eq:mp} imposes that $y_i=\neg x_i$.

Whenever $L=\emptyset$, it results that for any strict subset $K'\subsetneq K$, the transitions
generated from the sub-hypercube $\myh x {K'}$ form a (strict) subset of transitions generated
from $\myh x K$: the set of free components is a (strict) subset of $H$.
Therefore, it is useless to explore any subset of $K$.

Whenever $L\neq\emptyset$, by the $K$-closeness property of $\myh x K$, changing the state of a
component $i\in L$ would be \emph{irreversible} while updating only components in $K$.
Let us illustrate it by using the previous example without the $\vee x_3$ part: $f(x) = (1, x_1, \neg x_1 \wedge x_2)$.
With the initial configuration $x = 001$, one can obtain the sub-hypercube $\myh {001} {\{1,2,3\}} = ***$, $L = \{1\}$ and $J = \{2,3\}$.
$L = \{1\}$ means that component $1$ cannot return to its initial state when it has begun to flip.
Moreover, by the definition of MP transitions, all the components in $L$ are modified.
Therefore, whenever $L$ is not empty, one should consider all the sub-hypercubes $\myh x {K'}$ with
$(H\setminus L) \subseteq K' \subsetneq H$, to account for all these potential dependencies.
Let us consider $K' = H\setminus L = \{2,3\}$ in the previous example. One can obtain
$\myh {001} {\{2,3\}} = 00*$ and $L = \{3\}$, which allows discovering a new transition: ${001} \to_{\MP} {000}$.

This approach leads us to compute a set of sub-hypercubes $\myh x K$ from subsets $K$ of $\range n$, and where
each of them can be characterized by a triplet $(x,H,L)$ to which we refer to as a \emph{space}.
Each space $(x,H,L)$ characterizes a set of MP transitions from $x$
where the state of all the components in $L$ is flipped, as well as any subset of components in
$H\setminus L$:
the transitions $x\to_{\MP} y$ with $L\subseteq \Delta(x,y)\subseteq H$.
In our algorithm, the set of transitions generated by a space is never enumerated explicitly, as it is exponential in
$|H\setminus L|$. Instead, we count the number of transitions changing $m$ components, from $m=|L|$ (or $1$
whenever $L$ is empty) to $|H|$, i.e., the binomial coefficient $\binom{|H\setminus L|}{m-|L|}$.

Overall, the sampling of the configuration following $x$ can be summarized with the following steps:
\begin{enumerate}
    \item compute a set of spaces $S = \{(x,H^1,L^1), \ldots, (x,H^q,L^q)\}$, corresponding to the
        sub-hypercubes $\{\myh x {K^1}, \ldots, \myh x {K^q}\}$:
            starting with $K^1=\range n$,
            we consider the sub-hypercubes closed along $K^1\setminus \ell$ with each $\ell\subseteq L^1$,
        and recursively (thus whenever $L^1=\emptyset$, $q=1$).
    \item for each space, count the number of transitions it can generate per number of changing
        components;
    \item generate one random number to select the space $(x,H,L)$ and the number $m$ of components to flip;
    \item randomly select $m-|L|$ components in $H\setminus L$, let us denote by $C$ these chosen
        components;
    \item flip the state of the components in $L\cup C$.
\end{enumerate}
The size of sets $H$ and $L$ is at most $n$. However, the number of spaces $q$ can be exponential
with $n$ in the worst case.

Now, let us focus on the computation of $\myh x K$, the smallest sub-hypercube containing $x$ and
$K$-closed by $f$.
The main principle is to start from the sub-hypercube of rank $0$ with $x$ as the sole vertex, and
iteratively free components to fulfill the $K$-closure property.
To do so, we collect the set of components (among $K$) which can flip of state from at least one vertex
of the sub-hypercube: for each $i\in K$, if there exists a vertex $y$ of the sub-hypercube such that
$f_i(y)\neq x_i$, then $i$ must be free to verify the closeness property.
This process is then repeated until the $K$-closeness property is verified, which in the worst
case may require $n$ iterations.
It appears that the transitions generated by the asynchronous update mode match with the
components marked as free in the first iteration only: the components $i\in\range n$ such that
$f_i(x)\neq x_i$.
Thus, the additional transitions brought by the MP update mode are discovered in the later
iterations only.
This highlights the concept of \emph{permissive depth} we introduce in this paper:
the number of iterations in the computation of $\myh x K$ required to discover an MP transition.
A depth of 1 corresponds to the asynchronous transitions, while a depth of $n$ corresponds to the full MP
dynamics.
The simulation algorithm we propose in this paper allows controlling the depth of MP transitions,
for instance by following a probabilistic distribution.

\subsection{Algorithm}

Listings~\ref{lst:mp1} and \ref{lst:mp2} detail the steps for computing the reachable spaces and sampling the next
\begin{lstlisting}[float=t!,label={lst:mp1},caption={Computation of reachable spaces with the MP update mode}]
def can_flip(x: configuration, i: index, H: index set, v: bool):
    # assumes f is locally monotone
    if v == 1:
        z = min_configuration(x, i, H)
    else:
        z = max_configuration(x, i, H)
    return f[i](z) != v

def spread(x: configuration, K: index set, d: depth):
    # returns subset of K that can flip within the given depth
    H = {}
    repeat d times:
        H = H $\cup$ {i for i in K if can_flip(x, i, H, x[i])}
        K = K $\setminus$ H
    return H

def irreversible(x: configuration, H: index set):
    # returns subset of H that cannot flip back
    return {i for i in H if not can_flip(x, i, H, 1-x[i])}

def reachable_spaces(x: configuration):
    d = depth()
    S = [] # map of index set -> (index set, index set)
    K = {1, ..., n}
    Q = {K}
    while Q is not empty:
        K = Q.pop()
        H = spread(x, K, d) # H is subset of K
        L = irreversible(x, H) if d > 1 else {} # L is subset of H
        for each M non-empty subset of L:
            J = K \ M
            if J not in S and J not in Q:
                Q.push(J)
        S[K] = (H, L)
    return S
\end{lstlisting}
configurations from them, as sketched in the previous section.
In the description of the algorithms, we assumed fixed
(1) a BN $f$ of dimension $n$;
(2) a function \lstinline|depth| to determine the depth threshold for computing the sub-hypercubes (it has to be an
integer between 1 and $n$). For instance, it can be a constant, or a sampling from a discrete
distribution;
(3) a weighting vector $W\in\mathbb R_{\geq 0}^n$: $W_m$ is the weight of a transition modifying the
state of $m$ components simultaneously.
A uniform random walk along MP transitions is thus obtained with $W = \mathbf{1}_{n}$ and
\lstinline|depth=|$n$.

The \lstinline|spread| function computes the $K$-closure of the sub-hypercube starting from the
configuration $x$ and stops after $d$ iterations, $d$ being a given depth.
Determining whether the component $i$ can change state requires
determining the existence of a vertex $z$ of the sub-hypercube such that $f_i(z)\neq x_i$.
This is an instance of the classical Boolean satisfiability (SAT) problem, which, in general, is
NP-complete.
\begin{lstlisting}[float=*,label={lst:mp2},caption={Sampling of the next configurations with the MP update mode}]
def sample_next_configuration(x: configuration)
    S = reachable_spaces(x)
    if |S| = 0: return x   # fixed point
    # compute apparent rate of transitions
    R = $\mathbf 0$(len(S),n) # len(S)*n zero-filled matrix
    for i = 1 to len(S):
        if |L| > 0:
            R[i,|L|] = |L| * W[|L|]
        for j = 1 to |H$\setminus$ L|:
            R[i,|L|+j] = binom(|H$\setminus$L|,j) * W[|L|+j]
    r = $\mathcal U$[0, sum(R)[ # uniform sampling between 0 and sum(R) excluded
    s,m = where(cumsum(R) > r) # s = space, m = nb of components to flip
    H, L = S[s]
    C = L $\cup$ random.sample(H$\setminus$L, m - |L|)
    y = copy(x)
    y[C] = 1 - y[C]
    return y
\end{lstlisting}

The given algorithm makes the assumption that the BN $f$ is locally monotone.
In that case, given a sub-hypercube $h$, one can build in linear time from the influence graph $G(f)$
a configuration $x^{\mathrm{min}}\in \confs(h)$ so that $f_i(x^{\mathrm{min}}) = \min \{f_i(x)\mid
x\in \confs(h)\}$:
the idea is that for each component $j$ free in $h$, if $j$ has a positive (monotone) influence on
$i$ (i.e., $j\xrightarrow{+1}i \in G(f)$), then $x^{\mathrm{min}}_j = 0$;
if it has a negative influence on $i$, then $x^{\mathrm{min}}_j = 1$. If $j$ has no
influence on $i$, its state in $x^{\mathrm{min}}_j$ can be arbitrary.
A similar reasoning can be applied to compute
a configuration $x^{\mathrm{max}}\in \confs(h)$ so that $f_i(x^{\mathrm{max}}) = \max \{f_i(x)\mid x\in
\confs(h)\}$.
Then, one can decide in linear time whether $i$ can change state in $x$ in the scope of the
sub-hypercube $h$: if $x_i$ is 1, then $f_i(x^{\mathrm{min}})$ must be 0, and if $x_i$ is 0, then
$f_i(x^{\mathrm{max}})$ must be 1.
Without this assumption, the \lstinline|can_flip| function should be replaced with a call to a SAT solver.

\subsection{Correctness, complexity, and parametrization}
\label{sec:correctness}

The sampling of the next configuration is driven by two parameters: a distribution over permissive depth, and a weight $W \in\mathbb R^n_{\geq 0}$ for
transitions depending on the number of components they flip.
These parameters can affect the generated dynamics and the complexity of the sampling.

\paragraph{General case: full MP dynamics.}
Let us assume that the \lstinline|depth| function can always return $n$ (either it is a constant function
returning $n$, or the returned values follow a discrete distribution where the probability of
drawing $n$ is not 0), and that $W \in\mathbb R^n_{>0}$.
Then, any MP transition has a non-zero probability to be sampled.

Given a space $(x,H,L)$, as computed by \lstinline|reachable_spaces| and
characterized by a configuration $x$, a set $H\subseteq \range n$ of free components, and a subset
$L\subseteq H$ of irreversible components,
let us denote by $\tr(x,H,L)$ the set of candidate next configurations considered by
\lstinline|sample_next_configuration|:
\begin{equation}
    \tr(x,H,L) = \{ y\in \B^n\mid L\subseteq \Delta(x,y) \subseteq H\}\enspace.
\end{equation}
We prove that the set of transitions the algorithm can generate from the computed
spaces is equal to the full MP dynamics, except self-loops (Lemma~\ref{lem:complete}),
and that spaces generate disjoint sets of transitions (Lemma~\ref{lem:disjoint}).

\begin{lemma}\label{lem:complete}
    Given a BN $f$ of dimension $n$ and one of its configuration $x\in\B^n$,
    and denoting by $S$ the set of spaces returned by \lstinline|reachable_spaces|($x$) function,
    \[
        \bigcup_{(x,H,L)\in S} \tr(x,H,L) = \{ y \mid x\to_{\MP} y, x\neq y\}\enspace.
    \]
\end{lemma}

\begin{lemma}\label{lem:disjoint}
    Given a BN $f$ of dimension $n$ and one of its configuration $x\in\B^n$,
    and denoting by $S$ the set of spaces returned by \lstinline|reachable_spaces|($x$),
    for any distinct pair of spaces $(x,H,L)$, $(x,H',L')\in S$,
    $\tr(x,H,L)\cap \tr(x,H',L')=\emptyset$.
\end{lemma}

Proofs are given in Appendix~\ref{app:proofs}.

\paragraph{Worst-case complexity.}
Assuming locally-monotone BNs,
\lstinline|can_flip| is performed in linear time (we assume the influence graph $G(f)$ is given);
\lstinline|spread| makes in the worst case $n^2$ call to \lstinline|can_flip|, resulting in a cubic
time in $n$; whereas \lstinline|irreversible| is quadratic in $n$.
Function \lstinline|reachable_spaces| can then generate an exponential number of spaces.
The sampling is then linear with respect to the number of spaces generated, and linear with respect to $n$.
In the non-monotone case, \lstinline|can_flip| is an NP-complete problem which currently can be
solved in exponential time and space with SAT solvers.

\paragraph{Unitary depth: asynchronous and fully-asynchronous dynamics}
Let us consider the case whenever \lstinline|depth| function always returns 1.
The algorithm computes only one space $(x, H, L=\emptyset)$ with $H$ being the set of components $i$
such that $f_i(x)\neq x_i$, and can generate any transition to $y\neq x$ where $\Delta(x,y)\subseteq H$.
This corresponds exactly to the (general) asynchronous dynamics $\to_{\ga}$
assuming $W\in\mathbb R^n_{>0}$.
Moreover, as only one space is computed, the complexity drops to being linear in $n$, without any
assumption on the local-monotony of $f$.

Finally, whenever $W_1 > 0$ and for all $m \in\{2,\ldots,n\}$, $W_m = 0$, only transitions
modifying one component are generated, matching exactly with the fully-asynchronous dynamics
$\to_{\fa}$ with equiprobable transitions, and with complexity similar to the previous
restriction.

\subsection{Sampling reachable attractors}

The simulation of BNs is typically employed to assess the probability of reaching the different
attractors.
Because determining whether a configuration belongs to a cyclic attractor is a PSPACE-complete problem
with (a)synchronous update modes~\cite{MPBNs}, most simulation algorithms are parametrized with a
maximum number of steps to sample, without guarantee that an attractor has been reached.

In the case of the MP update mode, the attractors turn out to be exactly the smallest closed
sub-hypercubes (minimal trap spaces) of $f$, which can be computed at a much lower
cost, albeit still relying on SAT solving~\cite{MPBNs}.
Thus, instead of fixing an arbitrary number of simulation steps, one can verify during the simulation
whether the current configuration belongs to an attractor, and stop the sampling in that case.
(\lstinline|sample_reachable_attractor| of Listing~\ref{lst:attractor}).
\begin{lstlisting}[float=*,caption={Algorithms for sampling a reachable attractor with the MP update
mode},label={lst:attractor}]
def sample_reachable_attractor(x: configuration):
    stop = False
    while not stop:
        x = sample_next_configuration(x)
        every k iterations: # no need to verify at each step
            if in_attractor(x):
                stop = True
    A = reachable_attractors(x)
    return A[1] # A contains only one element

def filter_reachable_attractors(A: sub-hypercube list, x: configuration):
    H = spread(x, $\range n$, n)
    return [a for a in A if a $\preceq$ x/H] # a is smaller than the sub-hypercube
                                         # formed by x and H
def sample_reachable_attractor_bis(x: configuration):
    A = reachable_attractors(x) # list of attractors
    while len(A) > 1:
        x = sample_next_configuration(x)
        every k iterations: # no need to verify at each step
            A = filter_reachable_attractors(A, x)
    return A[1] # A contains only one element
\end{lstlisting}

In practice, the number of attractors reachable from a fixed initial configuration is usually small,
and can be efficiently enumerated with the MP update mode, for instance using the \textsc{mpbn}
tool~\cite{MPBNs}\footnote{\url{https://github.com/bnediction/mpbn}}.
In that case, the MP simulations can be employed to estimate the probability of reaching the
different attractors depending on the depth and weight parameters.
We then proceed as follows.
Before simulating, we first compute the full set of attractors reachable from the initial configuration $x$.
This set is then progressively refined during the simulation by removing attractors which are
no longer included in $\myh x {\range n}$. The simulation can then stop as soon as only one attractor
can be reached, i.e., the current configuration is in the strong basin of an attractor
(\lstinline|sample_reachable_attractor_bis| of Listing~\ref{lst:attractor}).

These two schemes assume that the full MP dynamics is sampled, or that the attractors of the sampled
dynamics match with the MP attractors as follows:
each MP attractor is a superset of one and only one attractor of the sampled dynamics, and each
attractor of the sampled dynamics is a subset of one and only one MP attractor.
This is always the case for BNs having no cyclic attractors.

\section{Evaluation}

Using a prototype written in
Python\footnote{\url{https://github.com/bnediction/mpbn-sim}},
we demonstrate the applicability of MP simulation on several Boolean models from the literature
comprising about thirty components.
After an illustration on toy examples, we study the effect of the parametrization on the assessment of
the robustness of mutations impacting the propensities of reachable attractors in those models.

Our MP simulation algorithm has two parameters: the permissive depth and the weight of
transitions $W$ depending on the number of binary state changes.
For the permissive depth, we will consider the constant $n$ (full MP dynamics), constant $1$
(asynchronous dynamics), and random sampling from a discrete exponentially decreasing distribution:
depth $d$ has probability $1/(2^d.M)$ with $M=\sum_{i=1}^n 1/2^i$ the normalization factor.
As depth $n$ has a non-zero probability, this parametrization also enables the full MP dynamics,
although largely prioritize transitions from low permissive depths.
Regarding $W$, we will consider the uniform weight $\mathbf 1_n$ (random walk),
and one-change only $1\mathbf 0_{n-1}$.

\subsection{Toy examples}

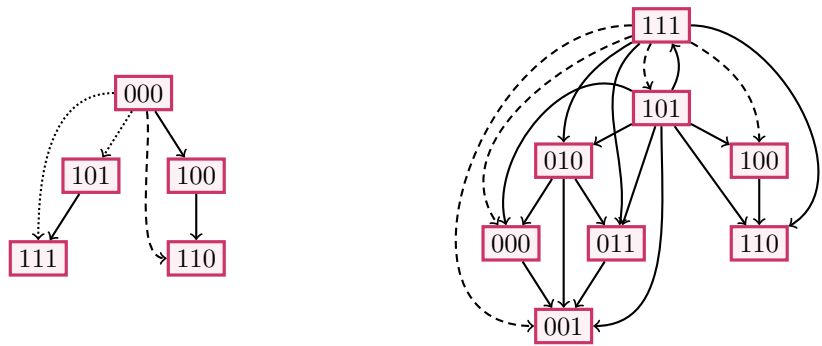
\begin{figure}[bt]
    \begin{minipage}{0.3\linewidth}
    \centering
    \begin{tikzpicture}[every text node part/.style={align=center}]

        \node[minimum width=0.60cm, draw, rectangle, color=purple!80, text=black!100, fill=purple!5, very thick] (3) at (0.7,-1.1) {$000$};
        \node[minimum width=0.60cm, draw, rectangle, color=purple!80, text=black!100, fill=purple!5, very thick] (5) at (0.0,-2.2) {$101$};
        \node[minimum width=0.60cm, draw, rectangle, color=purple!80, text=black!100, fill=purple!5, very thick] (6) at (1.4,-2.2) {$100$};
        \node[minimum width=0.60cm, draw, rectangle, color=purple!80, text=black!100, fill=purple!5, very thick] (7) at (-0.7,-3.3) {$111$};
        \node[minimum width=0.60cm, draw, rectangle, color=purple!80, text=black!100, fill=purple!5, very thick] (8) at (1.4,-3.3) {$110$};

        \draw[thick,densely dotted,->] (3) -- (5);
        \draw[thick,->] (3) -- (6);

        \draw[thick,->] (5) -- (7);

        \draw[thick,->] (6) -- (8);

        \draw[thick,densely dotted,->] (3) to [out=180,in=90] (7);
        \draw[thick,densely dashed,->] (3) to [out=280,in=180,looseness=0.6] (8);

    \end{tikzpicture}
\end{minipage}
\hfill
\begin{minipage}{0.6\linewidth}
    \centering
    \begin{tikzpicture}[every text node part/.style={align=center}]

        \node[minimum width=0.60cm, draw, rectangle, color=purple!80, text=black!100, fill=purple!5, very thick] (1) at (0.0,0.0) {$111$};
        \node[minimum width=0.60cm, draw, rectangle, color=purple!80, text=black!100, fill=purple!5, very thick] (2) at (0.0,-1.1) {$101$};
        \node[minimum width=0.60cm, draw, rectangle, color=purple!80, text=black!100, fill=purple!5, very thick] (3) at (-1.3,-1.8) {$010$};
        \node[minimum width=0.60cm, draw, rectangle, color=purple!80, text=black!100, fill=purple!5, very thick] (4) at (1.3,-1.8) {$100$};
        \node[minimum width=0.60cm, draw, rectangle, color=purple!80, text=black!100, fill=purple!5, very thick] (5) at (-2.0,-2.9) {$000$};
        \node[minimum width=0.60cm, draw, rectangle, color=purple!80, text=black!100, fill=purple!5, very thick] (6) at (-0.6,-2.9) {$011$};
        \node[minimum width=0.60cm, draw, rectangle, color=purple!80, text=black!100, fill=purple!5, very thick] (7) at (1.3,-2.9) {$110$};
        \node[minimum width=0.60cm, draw, rectangle, color=purple!80, text=black!100, fill=purple!5, very thick] (8) at (-1.3,-4.0) {$001$};

        \draw[thick,densely dashed,->] (1) to [out=240,in=120] (2);
        \draw[thick,->] (2) to [out=60,in=300] (1);
        \draw[thick,->] (1) to [out=210,in=90] (3);
        \draw[thick,densely dashed,->] (1) to [out=330,in=90] (4);
        \draw[thick,densely dashed,->] (1) to [out=200,in=125] (5);
        \draw[thick,->] (1) to [out=0,in=30] (7);
        \draw[thick,densely dashed,->] (1) to [out=180,in=180,looseness=1.25] (8);
        \draw[thick,->] (1) to [out=220,in=80] (6);

        \draw[thick,->] (3) -- (5);
        \draw[thick,->] (3) -- (6);
        \draw[thick,->] (3) -- (8);
        \draw[thick,->] (4) -- (7);
        \draw[thick,->] (5) -- (8);
        \draw[thick,->] (6) -- (8);

        \draw[thick,->] (2) -- (4);
        \draw[thick,->] (2) -- (7);

        \draw[thick,->] (2) -- (3);
        \draw[thick,->] (2) -- (6);

        \draw[thick,->] (2) to [out=150,in=105] (5);
        \draw[thick,->] (2) to [out=270,in=0] (8);

    \end{tikzpicture}
\end{minipage}

    \caption{MP transitions of BN~\eqref{eq:example1} from configuration $000$
        (left) and of BN~\eqref{ex:inputmodule} from configuration $111$ (right).
    Plain transitions have permissive depth $1$,
    dashed depth $2$, and dotted depth $3$.%
    \label{fig:MPToys}}
\end{figure}

Let us first consider the bi-stable example $\exone$ from the configuration $000$.
This BN has two fixed points: $110$ and $111$.
As explained in Sect.~1, the (a)synchronous dynamics from $000$ predicts only one trajectory:
$000\to_{\ga} 100\to_{\ga} 110$.
Thus, only one attractor is reachable (100\% propensity).
The MP dynamics uncovers another reachable attractor ($111$), as depicted in
Fig.~\ref{fig:MPToys}(left).
Considering all the transitions, the MP simulation would conclude that the reachability of
these two attractors are equiprobable: from the 4 initial transitions, 2 lead to the strong basin
of $110$ and two to the strong basin of $111$.
Whenever the depth is a random variable with an exponentially decreasing distribution, reaching
$111$ requires drawing a depth of $3$ (probability of 1/7), leading to reaching
$111$ with probability 1/14, and $110$ with probability 13/14.
This indicates the sensitivity of the reachability of $111$ to the permissive depth, and thus to the
time scales of the underlying quantitative model.

Let us now consider the example $\extwo$ where the two attractors reachable with the asynchronous mode
(Fig.~\ref{fig:inputmodule}(c)) are identical to ones reachable with MP mode (Fig.~\ref{fig:MPToys}(right)).
In fully-asynchronous mode, they have an equal propensity, however in the (general) asynchronous
mode, $001$ propensity is twice the one of $110$.
Whereas the MP dynamics uncover additional trajectories and reachable configurations, in this
particular example, the propensities of attractors is the same as with the general asynchronous
case.

An important aspect of MP dynamics is that it is transitive: any state reachable in a sequence of
transitions is reachable in one transition.
Thus, the size of the (reachable) strong basins of attractors can contribute significantly to their
propensities.
In the above two examples, the propensities of the attractors in MP uniform random walk is entirely
determined by the size of their basins.

It should be also stressed that the parameters for determining the transition probabilities are
usually arbitrary.
Thus, analyzing the impact of parameters on the probability of reaching the
different attractors by a random walk in the MP dynamics may only give an empirical insight on
the Boolean dynamics, and cannot be formally transfered to an associated quantitative model.

\subsection{Models from literature with different mutation conditions}

From the literature, we selected BNs modeling cell fate decision processes:
the reduced cell death receptor model~\cite{CellDeath2010} with 14 components;
the tumor invasion model~\cite{TumorInvasion} with 32 components; and
the bladder model~\cite{Bladder2015} with 35 components.
These models have been designed with the fully-asynchronous update mode, and have been evaluated
with respect to their ability to predict the changes of the attractors propensities subject to
different mutation conditions (modeled by forcing some components to some state).
For each model, we performed 10,000 simulations from the relevant initial configurations
and for several mutation conditions, with different simulation parameters.
With our prototype, the computational cost of the permissive depth is substantial, the simulations being between 3 times
to 50 times slower than with depth 1 only, depending on the number of reachable spaces computed at
each simulation step.
However, no particular optimization has been implemented.

\begin{figure}[bt]
    \centering
    \includegraphics[width=\textwidth]{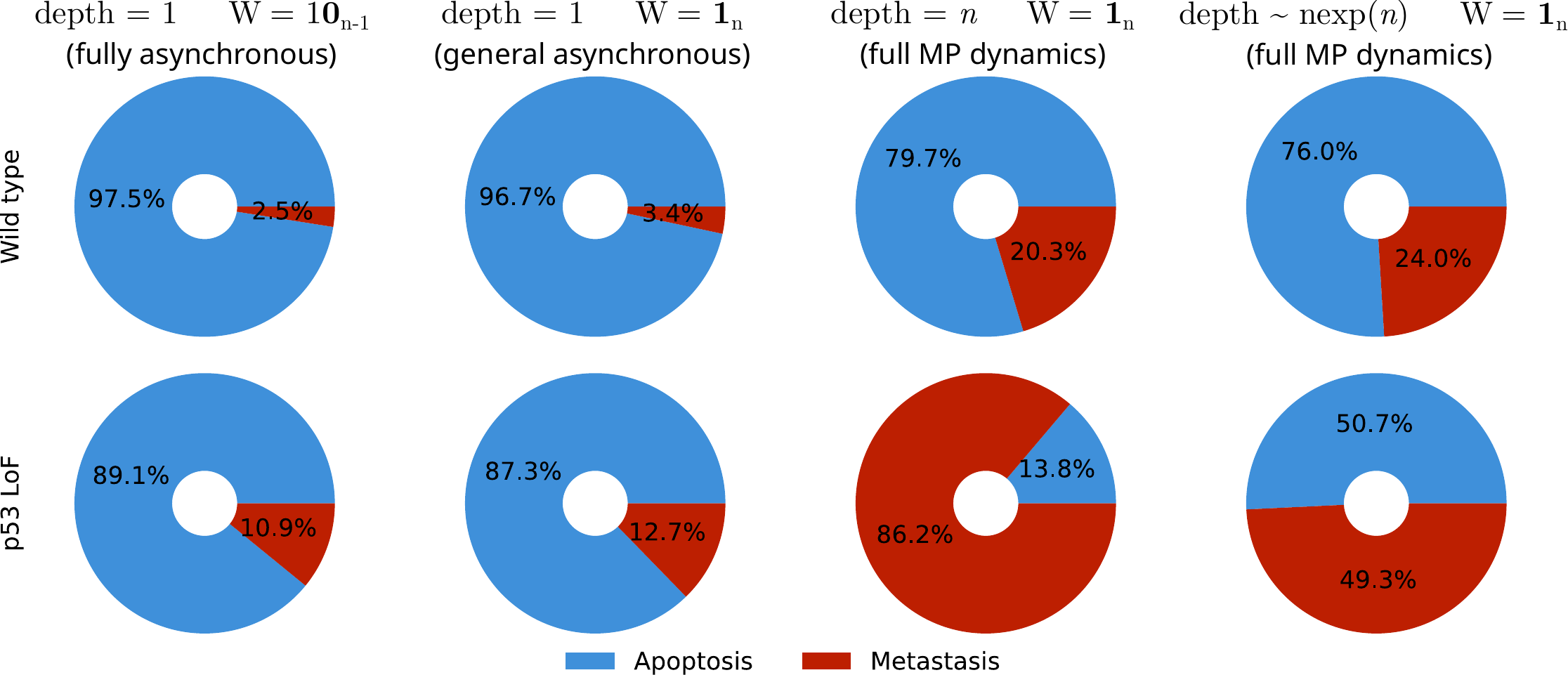}
    \caption{Estimated propensities of reachable attractors from a unique configuration of the tumor
    invasion model in two mutation conditions (no mutation; p53 forced to 0), and different
simulation parameters.\label{fig:tumor}}
\end{figure}

Fig.~\ref{fig:tumor} gives an example of the estimated propensities of attractors reachable from a
unique initial configuration of the tumor invasion model with different mutation conditions and parameters.
On the one hand, we can observe that the permissive depth can change drastically the proportions of reachable
attractors: there are much more paths to the Metastasis attractor when considering potential time scale
heterogeneity.
The difference can be reduced by still considering the full MP dynamics, but giving
exponentially decreasing probability to transitions with high permissive depth
(computing transitions at depth at least $D$ is done with probability
$\sum_{d=D}^n (2^d\cdot\sum_{i=1}^n 2^{-i})^{-1}$, with $n$ the number of nodes in the model, 32 in
this case).
On the other hand, the qualitative effect of the p53 mutation is similar whenever analyzed only with
the asynchronous simulations (permissive depth = 1), or in the most permissive cases: the propensity
of the Metastasis attractor is larger in the mutant.

From a global analysis on the 3 models\footnote{See \url{https://doi.org/10.5281/zenodo.6725844}
for supplementary information},
we observe that the absolute value of the propensities of the reachable attractors
changes substantially when considering MP dynamics, in particular with depth fixed to $n$ and
uniform transition rates.
Variable depth enables reducing the difference with the asynchronous dynamics  while giving access to the full MP dynamics.
Regarding the effect of mutations, it appears that, in most cases, the propensities of reachable
attractors are affected in qualitatively the same direction, with various MP simulation parameters.
This may indicate that these perturbations should be robust to heterogeneous time scales in the
quantitative models captured by the BN abstraction.
On the contrary, perturbations having a different qualitative effect on attractor propensities when considering
permissive depth may be sensitive to the variety of time scales of the actual system, which could not be
captured with usual (a)synchronous modes.

\section{Discussion}

The simulation of (asynchronous) BNs is a usual task in systems biology applications for
assessing the effect of genetic perturbations on the propensities of the different cellular phenotypes.
However, it is known that the asynchronous dynamics of BNs can preclude behaviors observed in
quantitative systems, which may lead to substantial biases for the aforementioned analysis.
On the contrary, the MP update mode ensures the completeness of the Boolean dynamics.

In this paper, we provide a first algorithm to sample trajectories from MP dynamics.
The additional transitions predicted by MP come from iterative computations of possible state
changes for each component, and where the first iteration corresponds to the asynchronous updates.
By parameterizing the \emph{depth} of this computation, we obtain a generalization of transition
sampling from BNs with MP, (general) asynchronous, and fully-asynchronous dynamics.
The main bottleneck of the provided algorithm is its potential exponential blow-up when
considering all reachable spaces.
Further work will address its efficient implementations and approximations.

As illustrated on different models from the literature, the simulation with the MP mode can largely affect
predictions, both in terms of reachable attractors, and in terms of their propensities.
Because of the transitive properties of MP dynamics (there is a direct transition to any reachable
configuration), the attractors having a large (reachable) strong basin will dominate an equiprobable
random walk of MP dynamics.
Variable-depth simulation can then give more weight to transition with low permissive depth and
sooth the effect of the size of the basins while still capturing the full MP dynamics.

It should be noted that, as with usual simulations of asynchronous BNs~\cite{Mendes2018,Stoll2012},
the estimated attractor propensities are purely empirical and do not relate formally to the actual
propensities in the modeled quantitative system.
Due to its abstraction level, a BN is intrinsically non-deterministic. Assigning
probabilities to transitions is a very strong assumption which, in this context, cannot be
justified with any modeling or physical principle. At this level of abstraction, the behavior of
the modeled system is not a Markov process and cannot be approximated by a Markov process
(would the coefficient be fitted on data, for instance).

From a modeling perspective, our algorithm could be extended to have the permissive depth
being different among components, enabling to fine-tune the time scale of their updates: the
slower components should allow the more permissive depth.
This would bring further control over the transitions added by the MP mode, while capturing
trajectories precluded by the asynchronous mode.

\subsubsection{Acknowledgments}
This work was supported by the French Agence Nationale pour la Recherche (ANR),
with the project ``\href{https://bnediction.github.io}{BNeDiction}'' (ANR-20-CE45-0001).

\appendix

\section{Proofs}
\label{app:proofs}

\subsection{Proof of Lemma 1}
Consider any $(x,H,L)\in S$.
Let $h\in\Hd^n$ be the sub-hypercube where dimensions in $H$ are free, and otherwise fixed as in
$x$: for each $i\in\range n$, if $i\in H$, then $h_i=*$, otherwise, $h_i=x_i$.
In such a configuration $y$, for each $i\in H$, there exists a
configuration  $z\in\confs(h)$ such that $f_i(z)=y_i$:
indeed, if $y_i=x_i$, then $i\notin L$.
Thus, $x\to_{\MP} y$ by instantiating Eq.~\eqref{eq:mp} with $K=H$.

Conversely, let us assume that there exists $K\subseteq\range n$ leading to $x\to_{\MP} y$, and let
$H^0$ be the set of free components in $\myh x K$ (note that $\Delta(x,y)\subseteq H^0$).
Let $(x, H^1, L^1)$ be the first space computed by our algorithm (i.e., from $K=\range n$).
By construction, $H^1$ is the set of free components in $\myh x {\range n}$, thus
$H^0\subseteq H^1$.
Two cases arise:
either $L^1\subseteq\Delta(x,y)$ and then the transition to $y$ is generated from this first space,
or there exists at least one component $i\in L^1$ whereas $x_i=y_i$.
Recall that a component $i$ can be in $L^1$ only if from each vertex $z$ of the sub-hypercube
$f_i(z)\neq x_i$.
Thus $K$ is necessarily a strict subset of $H^1$, excluding at least the
component $i$.
Therefore, $H^0 \subseteq K \subseteq H^1\setminus\{ i\in L^1\mid x_i=y_i\}=H^2$,
and by construction, $(x,H^2, L^2)\in S$ with $L^2\subseteq\Delta(x,y)$.

\subsection{Proof of Lemma 2}
\label{sec:proof-disjoint}

The set of spaces computed by \lstinline|reachable_spaces| can be inductively characterized by
a map $S^l$  for some $l\in\mathbb N$
from sets of components $K$ to their associated sub-hypercube $\myh x K$ characterized by $(H,L)$,
as defined below ($x$ being fixed it is omitted):
\begin{align}
    S^1 &= \{ \range n \mapsto (H^0,L^0) \} \\
    S^{k+1} &= S^k \cup \{ K \setminus M \mapsto (H',L') \mid K \mapsto (H,L)\in S^i,
    \emptyset \subsetneq M \subseteq L \}
\end{align}
Recall that $\forall K\mapsto (H,L)\in S^k$, $L\subseteq H\subseteq K$,
and $\tr(x,H,L) = \{ y\in\B^n\mid L\subseteq \Delta(x,y)\subseteq H\}$.

\noindent
We prove that for any $k\in\mathbb N$, $\forall K\mapsto (H,L),K'\mapsto (H',L')\in S^k, K\neq K'$,
\[\tr(x,H,L)\cap\tr(x,H',L')=\emptyset\enspace.\]

Let us first consider the cases whenever
$L\not\subseteq H'$ or $L'\not\subseteq H$:
by $\tr$ definition, $\tr(x,H,L)\cap \tr(x,H',L')=\emptyset$.
Indeed, in the first case, remark that $\forall y\in\tr(x,H,L)$,  $L\setminus H'\subseteq
\Delta(x,y)$ while $\forall y'\in\tr(x,H',L')$, $(L\setminus H')\cap\Delta(x,y')=\emptyset$,
thus $y\neq y'$. The second case is a symmetry.

We establish the following propositions:
\begin{itemize}
\item (P1) Any component $i\in\range n$ such that there exists $K\mapsto (H,L)$ with $i\in L$
    verifies $f_i(x)=\neg x_i$.
    Thus, $\forall K'\mapsto (H',L')$, $i\in K'\implies i\in H'$.

\item
(P2) By definition of $S^k$, $K \mapsto (H,L)\in S^k \implies \forall i \in \range n\setminus K$
 there exists $K'\mapsto (H',L')\in S^{k-1}$ with $i\in L'$.
 Thus by P1, $f_i(x)=\neg x_i$.

\item
(P3) By sub-hypercube $K$-closeness definition and minimality,\\
$\forall K\mapsto (H,L), K'\mapsto (H',L')\in S^k, K'\subseteq K \implies
K'\cap L\subseteq L'$ $\subseteq H'$.

\item
(P4) for any $K\neq\range n$, $K\mapsto (H,L)\in S^k\setminus S^{k-1}\implies
\exists K'\mapsto (H',L')\in S^{k-1}$ with $K\subsetneq K'$ and $K'\setminus K\subseteq L'$
(by P2).
\end{itemize}

Consider any $K\mapsto (H,L),K'\mapsto (H',L')\in S^k, K\neq K'$, such that both $L\subseteq H'$ and
$L'\subseteq H'$.
By P4, there exists $K''\mapsto (H'',L'')\in S^{k-1}$ with $(K \cup K')\subseteq K''$ and such that
there exists $i\in L''$ where $i\notin K$
(note: it always work with $K''=\range n$).
If $i\in K'$, then $i\in L'$ (by P3), thus $L'\not\subseteq H$, a contradiction.
Thus, $i\notin K'$.
By induction using P2 and P4, we obtain that $K'\subseteq K'$.
By symmetry (apply the same reasoning by swapping $K$ and $K'$), $K=K'$.

\bibliographystyle{splncs04}
\bibliography{cmsb22}

\begin{thebibliography}{10}
\providecommand{\url}[1]{\texttt{#1}}
\providecommand{\urlprefix}{URL }
\providecommand{\doi}[1]{https://doi.org/#1}

\bibitem{CellDeath2010}
Calzone, L., Tournier, L., Fourquet, S., Thieffry, D., Zhivotovsky, B.,
  Barillot, E., Zinovyev, A.: Mathematical modelling of cell-fate decision in
  response to death receptor engagement. PLOS Computational Biology
  \textbf{6}(3),  e1000702 (2010). \doi{10.1371/journal.pcbi.1000702}

\bibitem{TumorInvasion}
Cohen, D.P.A., Martignetti, L., Robine, S., Barillot, E., Zinovyev, A.,
  Calzone, L.: Mathematical modelling of molecular pathways enabling tumour
  cell invasion and migration. PLOS Computational Biology  \textbf{11}(11),
  e1004571 (2015). \doi{10.1371/journal.pcbi.1004571}

\bibitem{Crama2011}
Crama, Y., Hammer, P.L.: Boolean Functions. Cambridge University Press (2011).
  \doi{10.1017/cbo9780511852008}

\bibitem{Ishihara2005}
Ishihara, S., Fujimoto, K., Shibata, T.: Cross talking of network motifs in
  gene regulation that generates temporal pulses and spatial stripes. Genes to
  Cells  \textbf{10}(11),  1025--1038 (2005).
  \doi{10.1111/j.1365-2443.2005.00897.x}

\bibitem{kauffman69}
Kauffman, S.A.: Metabolic stability and epigenesis in randomly connected nets.
  Journal of Theoretical Biology  \textbf{22},  437--467 (1969).
  \doi{10.1016/0022-5193(69)90015-0}

\bibitem{Mendes2018}
Mendes, N.D., Henriques, R., Remy, E., Carneiro, J., Monteiro, P.T., Chaouiya,
  C.: Estimating attractor reachability in asynchronous logical models.
  Frontiers in Physiology  \textbf{9} (2018). \doi{10.3389/fphys.2018.01161}

\bibitem{Montagud22}
Montagud, A., B{\'{e}}al, J., Tobalina, L., Traynard, P., Subramanian, V.,
  Szalai, B., Alföldi, R., Pusk{\'{a}}s, L., Valencia, A., Barillot, E.,
  Saez-Rodriguez, J., Calzone, L.: Patient-specific boolean models of
  signalling networks guide personalised treatments. {eLife}  \textbf{11}
  (2022). \doi{10.7554/elife.72626}

\bibitem{MPBNs}
Paulev{\'e}, L., Kol{\v c}{\'a}k, J., Chatain, T., Haar, S.: Reconciling
  qualitative, abstract, and scalable modeling of biological networks. Nature
  Communications  \textbf{11}(1) (2020). \doi{10.1038/s41467-020-18112-5}

\bibitem{automata21}
Paulev{\'e}, L., Sen{\'e}, S.: {Non-deterministic Updates of Boolean Networks}.
  In: 27th IFIP WG 1.5 International Workshop on Cellular Automata and Discrete
  Complex Systems (AUTOMATA 2021). Open Access Series in Informatics (OASIcs),
  vol.~90, pp. 10:1--10:16. Schloss Dagstuhl -- Leibniz-Zentrum f{\"u}r
  Informatik, Dagstuhl, Germany (2021). \doi{10.4230/OASIcs.AUTOMATA.2021.10}

\bibitem{VLBNs}
Paulevé, L.: {VLBNs - Very Large Boolean Networks} (2020),
  \url{https://doi.org/10.5281/zenodo.3714876}

\bibitem{Bladder2015}
Remy, E., Rebouissou, S., Chaouiya, C., Zinovyev, A., Radvanyi, F., Calzone,
  L.: A modeling approach to explain mutually exclusive and co-occurring
  genetic alterations in bladder tumorigenesis. Cancer Research
  \textbf{75}(19),  4042--4052 (2015). \doi{10.1158/0008-5472.can-15-0602}

\bibitem{Rodrigo2011}
Rodrigo, G., Elena, S.F.: Structural discrimination of robustness in
  transcriptional feedforward loops for pattern formation. {PLOS} {ONE}
  \textbf{6}(2),  e16904 (2011). \doi{10.1371/journal.pone.0016904}

\bibitem{Schaerli2014}
Schaerli, Y., Munteanu, A., Gili, M., Cotterell, J., Sharpe, J., Isalan, M.: A
  unified design space of synthetic stripe-forming networks. Nature
  Communications  \textbf{5}(1) (2014). \doi{10.1038/ncomms5905}

\bibitem{Stoll2012}
Stoll, G., Viara, E., Barillot, E., Calzone, L.: Continuous time boolean
  modeling for biological signaling: Application of gillespie algorithm. {BMC}
  Systems Biology  \textbf{6}(1), ~116 (2012). \doi{10.1186/1752-0509-6-116}

\bibitem{thomas73}
Thomas, R.: Boolean formalization of genetic control circuits. Journal of
  Theoretical Biology  \textbf{42}(3),  563 -- 585 (1973).
  \doi{10.1016/0022-5193(73)90247-6}

\bibitem{Zanudo21}
Za{\~{n}}udo, J.G.T., Mao, P., Alcon, C., Kowalski, K., Johnson, G.N., Xu, G.,
  Baselga, J., Scaltriti, M., Letai, A., Montero, J., Albert, R., Wagle, N.:
  Cell line-specific network models of {ER}$+$ breast cancer identify potential
  {PI}3k$\alpha$ inhibitor resistance mechanisms and drug combinations. Cancer
  Research  \textbf{81}(17),  4603--4617 (2021).
  \doi{10.1158/0008-5472.can-21-1208}

\end{thebibliography}

\end{document}